# A domain-specific atomistic model for Earth's core materials

Tianqi Wan[1#], Liangrui Wei[1#], Zepeng Wu[1], Renata M. Wentzcovitch[2-4], Yang Sun[1*]

[1]*Department of Physics, Xiamen University, Xiamen, 361005, China*

[2]*Department of Applied Physics and Applied Mathematics, Columbia University, New York, NY, 10027, USA*

[3]*Department of Earth and Environmental Sciences, Columbia University, New York, NY, 10027, USA*

[4]*Lamont–Doherty Earth Observatory, Columbia University, Palisades, NY, 10964, USA*

Understanding Earth's core requires accurate descriptions of multicomponent iron alloys under extreme conditions. Here, we develop the Earth's Core Atomistic Model (ECAM) for Fe–Ni–S–Si–O–C–H alloys through sampling with a foundational atomistic model, targeted *ab initio* simulations, and iterative exploration. A compact dataset selected from nearly half a million configurations is consistently labeled using high-accuracy first-principles calculations and used to train a lightweight model. The model explicitly includes thermal electronic contributions and runs nearly two orders of magnitude faster than the original foundational model. ECAM corrects artifacts in the foundational model's description of core materials and reproduces the density, seismic velocity, Gibbs free energy, and phase diagrams of Fe and binary alloys in agreement with previous *ab initio* calculations. ECAM is expected to enable more realistic modeling of chemical partitioning and crystallization in Earth's core. Our workflow also provides a practical route from foundational-model sampling to efficient domain-specific simulation.

## I. Introduction

Constraining the core's composition is essential for understanding the evolution of the deep Earth and the processes that sustain Earth's magnetic field[1–3]. Earth's core consists predominantly of Fe alloyed with Ni and lighter elements such as Si, O, S, C, and H[4,5]. The concentrations of these elements are constrained by the seismic properties and solid–liquid partitioning behavior of multicomponent alloys[6–10]. Given the core's extremely high pressures and temperatures, limited laboratory experiments must be combined with theoretical modeling across the complex septenary chemical space of core-forming materials to satisfy seismic and geochemical constraints[11–21].

*Ab initio* simulations have been the primary approach for calculating the seismic properties, partitioning behavior, and phase diagrams of core materials[16,20,22–28]. While these calculations have provided key information for current models of core composition, most studies have relied on combining or extrapolating results from binary or ternary iron alloys to describe multicomponent systems[29]. However, complex alloying effects can alter material properties compared with those of the corresponding binary and ternary alloys[30]. The high computational cost of *ab initio* simulations severely limits the study of the septenary Fe–Ni–S–Si–O–C–H system[31].

Machine-learning interatomic potentials (MLIPs) can reduce the computational cost by learning energies and forces from *ab initio* calculations, enabling simulations of larger systems over longer timescales[32–34]. Customized MLIPs have been shown to provide satisfactory descriptions of pure Fe and a few binary and ternary iron alloys under core conditions, including their melting behavior and phase competition[30,35–40]. Nevertheless, constructing models across the septenary compositional space of core materials requires large-scale configurational sampling and *ab initio* labeling, which remain computationally demanding. Recently developed foundational atomistic models (FAMs) offer a way to accelerate this exploration[41–44]. Their chemically diverse training datasets support sampling across many elements and structures. However, broad chemical coverage does not ensure accurate interactions in different alloys, especially under the core's extreme conditions[45]. Our recent assessment of 17 FAMs suggests that none of the tested models consistently reproduces all first-principles properties of core materials. The main limitations are inaccurate descriptions of atomic interactions under extreme conditions and the omission of electronic entropy contributions[31].

Although fine-tuning FAMs can improve their accuracy for specific systems[46], parameter updates alone leave the model's input variables and computational architecture unchanged. Simulations of Earth's core materials require a free energy surface with electronic entropy contributions to accurately capture the small free-energy differences governing phase competition. Errors in the foundational model can also bias configurational sampling and leave essential atomic environments poorly represented. The central challenge is therefore to combine broad FAM exploration with targeted first-principles sampling and labeling to train an independent, computationally efficient free energy model for large-scale dynamics. Here, we develop the Earth's Core Atomistic Model (ECAM) for the septenary Fe–Ni–S–Si–O–C–H system, focusing on pressure–temperature conditions between the inner-core boundary and Earth's center. We perform broad configurational sampling using the pretrained MatterSim model [43]. To address artifacts in MatterSim's description of specific atomic environments, we design targeted *ab initio* sampling to complement its broad configurational exploration. From the combined sampling, we construct a compact dataset that represents the relevant atomic environments and enables efficient training of lightweight neural networks. We validate ECAM by comparing its predictions of density, seismic velocity, and Gibbs free energy with results from previous *ab initio* calculations. Our workflow illustrates a route from broad foundational-model sampling to an efficient domain-specific model, providing a basis for future studies of light-element partitioning and crystallization in multicomponent alloys under core conditions.

[#]Equal contribution.
[*]Email: yangsun@xmu.edu.cn

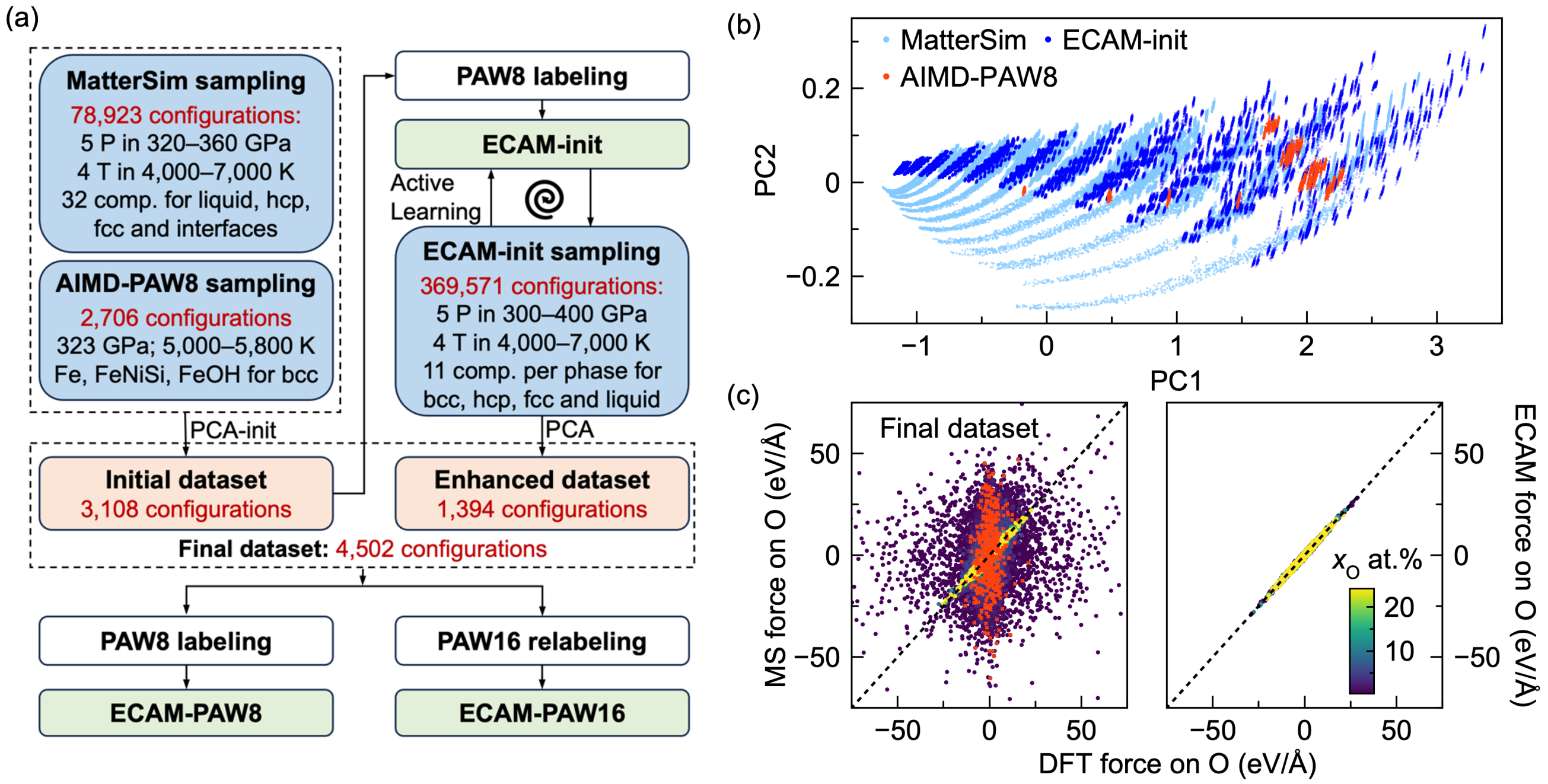

**FIG. 1 Data collection and the ECAM training.** (a) Workflow of configurational sampling, dataset preparation, DFT labeling, and model training. (b) PCA projections of configurations sampled by MatterSim, AIMD-PAW8, and ECAM-init. (c) Oxygen force components predicted by MatterSim (left) and ECAM-PAW8 (right), compared with DFT-PAW8 reference values for configurations in the final dataset. Points are colored by the oxygen concentration, $x_O$ (at.%). Red crosses identify force components from configurations sampled by AIMD-PAW8. Dashed lines indicate perfect agreement.

## II. Results

### 1. Training and validation

To generate the training dataset, we first perform configurational sampling using MD driven by the pretrained MatterSim model. We sample five pressures from 320 to 360 GPa at intervals of 10 GPa and four temperatures from 4000 to 7000 K at intervals of 1000 K. To cover plausible core compositions, we generate 11 compositions by interpolating between pure Fe and the septenary endpoint $Fe_{35}Ni_{10}S_5Si_{10}O_{15}C_{10}H_{15}$. This endpoint encompasses the upper bounds of the Ni and light-element concentrations in Earth's core estimated by [9]. The interpolated composition is therefore $Fe_{100-65\lambda}Ni_{10\lambda}S_{5\lambda}Si_{10\lambda}O_{15\lambda}C_{10\lambda}H_{15\lambda}$, where λ ranges from 0 to 1 in increments of 0.1. We generate these compositions in the bcc, hcp, fcc, and liquid phases using ~ 250 atoms per cell. Consistent with our previous assessment[31], MatterSim frequently produces artificial O aggregation at low O concentrations, resulting in many unphysical O clusters in the sampled configurations. To prevent this aggregation, we replace all but one O atom with Fe in each system before systematic sampling. Since single-O configurations do not provide information on O–O interactions, we add one O-rich liquid ($Fe_{150}Ni_{15}Si_{25}S_{15}H_{30}O_{80}C_{15}$) for which MatterSim was found to perform reliably [31]. We thus sample 32 composition–phase combinations. During the solid-phase simulations, we observe that C, H, and O spontaneously exhibit superionic behavior at higher temperatures, consistent with previous AIMD simulations [16]. Because MatterSim substantially underestimates the bcc melting point[31], many initially bcc structures spontaneously transform to hcp or liquid. We therefore supplement the bcc dataset with AIMD-PAW8 trajectories for pure Fe and Fe–Ni–Si systems. We also use AIMD to extend sampling of O-poor Fe–O–H systems with O concentrations of 0.4, 0.8, and 1.1 at.% at 323 GPa and 5 temperatures between 5000 and 5800 K.

MatterSim and AIMD sampling generate 81,629 atomic configurations. Because configurations sampled directly from MD trajectories can be strongly correlated, we use PCA-based selection to reduce redundancy in the dataset. We divide the PCA map into a uniform grid and select 3,108 representative structures from the grid cells to form the initial dataset. As shown in Fig. 1(b), MatterSim sampling covers a broad region of the PCA map, whereas AIMD sampling is sparser and more localized. However, for the dilute-O configurations sampled by AIMD, MatterSim shows substantial deviations from the DFT reference forces on O atoms, as shown in Fig. 1(c). These AIMD configurations are therefore critical for improving the coverage of dilute-O environments in the initial dataset.

The configurations in the initial dataset are labeled using DFT calculations with the PAW8 settings to obtain energies, forces, and virials. During labeling, we identify a small number of MatterSim-sampled configurations with anomalously large DFT atomic forces (>100 eV/Å). These configurations contain locally unphysical atomic arrangements that can bias subsequent training toward irrelevant atomic environments. Thus, we relax them through additional AIMD simulations. After only 10 AIMD steps, the large forces decrease without substantial changes to the overall atomic structures. We replace the original configurations with these relaxed structures and find an improvement in training accuracy. Simulations with ECAM-init already show no unphysical short O–O contacts in the O-poor liquid.

Using ECAM-init as the starting model, we continue sampling the $Fe_{100-65\lambda}Ni_{10\lambda}S_{5\lambda}Si_{10\lambda}O_{15\lambda}C_{10\lambda}H_{15\lambda}$ compositions in the bcc, hcp, fcc, and liquid phases while iteratively retraining the model. This active-learning process samples additional configurations containing multiple O atoms that MatterSim cannot reliably access. In total, the sampling generates 369,571 configurations, which are analyzed together with the initial dataset using PCA. As shown in Fig. 1(b), the additional configurations sampled with ECAM-init fill gaps in the initial MatterSim-generated distribution and extend its coverage. PCA selection yields 1,394 additional configurations, forming the enhanced dataset. Combining the initial and enhanced datasets yields a final dataset spanning diverse chemical compositions and phases. Further analysis of the PCA distribution in Supplementary Fig. S1 suggests that the distinct bands reflect differences in Fe content. The hcp, bcc, and fcc configurations occupy nearby regions, suggesting similarities in their local atomic environments, whereas the liquid configurations span a broader region. Detailed configurational information on the final dataset is summarized in Supplementary Table S1.

ECAM-PAW8 is trained on the final dataset labeled using DFT calculations with the PAW8 settings. As shown in Fig. 1(c), ECAM-PAW8 substantially reduces the errors in MatterSim-predicted forces on O atoms and closely reproduces the DFT reference values. We also relabel all configurations in the final dataset using the more accurate PAW16 settings, which treat additional inner-shell electrons as valence electrons and use a higher plane-wave kinetic-energy cutoff and denser k-point sampling (see Methods). Previous studies suggest that PAW16 settings substantially improve the accuracy of free energies and melting temperatures for iron under inner-core conditions[24,25,38]. We then train ECAM-PAW16 on the relabeled dataset. The model closely reproduces the DFT training data, with root-mean-square errors of 4.9 meV/atom for energies and 0.332 eV/Å for forces (Supplementary Fig. S2).

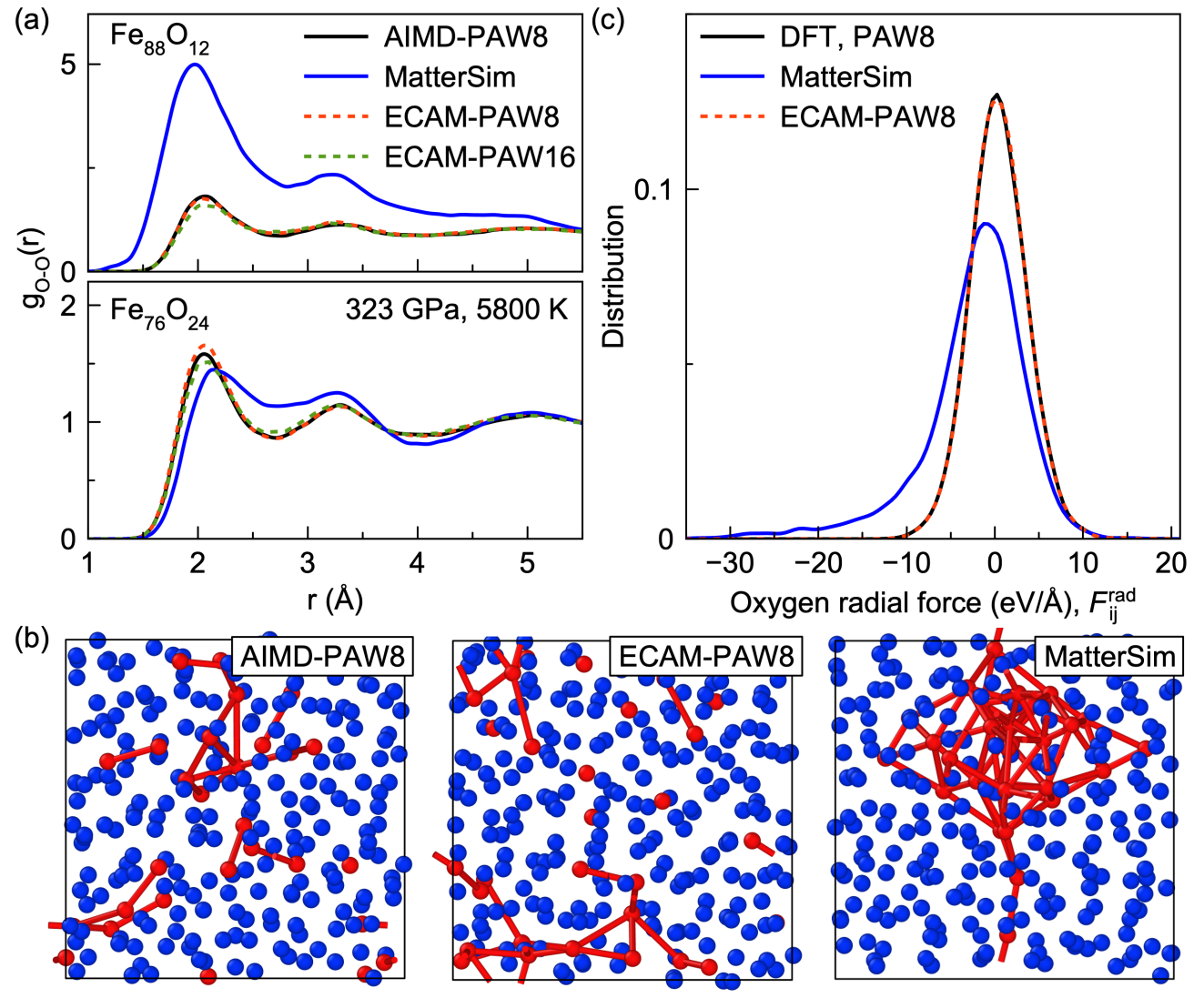

**FIG. 2 Resolved artificial oxygen aggregation.** (a) O–O partial PCF, g(r), for $Fe_{88}O_{12}$ and $Fe_{76}O_{24}$ liquids at 323 GPa and 5800 K, calculated using AIMD and different MLIPs. (b) Snapshots of the $Fe_{88}O_{12}$ liquid from AIMD-PAW8, ECAM-PAW8, and MatterSim simulations. (c) O–O radial force distributions for configurations in the final dataset, evaluated using DFT and MLIP forces. The radial force is the projection of the relative force between two O atoms onto their interatomic displacement direction. For an O–O pair consisting of atoms i and j, the radial force is defined as $F_{ij}^{\mathrm{rad}} = \left(f_j - f_i\right) \cdot \frac{\mathrm{r}_{ij}}{|\mathrm{r}_{ij}|}$, where $f_i$ and $f_j$ denote the forces acting on atoms i and j, respectively, and $\mathrm{r}_{ij}$ denotes their displacement vector. The distributions include only nearest-neighbor O–O pairs within 4 Å. A negative radial force indicates an attractive tendency to decrease the O–O distance.

We first test whether ECAM resolves the artificial O aggregation observed in MatterSim simulations of O-poor liquids[31]. Figure 2(a) compares the O–O partial pair correlation functions (PCF) for Fe–O liquids containing 12 and 24 at.% O at 323 GPa and 5800 K. At 12 at.% O, MatterSim produces an anomalously high first peak near 2 Å, indicating substantially overestimated short-range O–O correlations. This discrepancy persists at 24 at.% O but is less pronounced. The MatterSim snapshot in Fig. 2(b) exhibits an extended O aggregate, whereas O atoms remain dispersed in the AIMD simulation. Both ECAM-PAW8 and ECAM-PAW16 closely reproduce the AIMD pair correlations and show no artificial O aggregation. To identify the origin of this aggregation, we compare the radial-force distributions for nearest-neighbor O–O pairs evaluated on the same configurations in Fig. 2(c). The ECAM-PAW8 distribution closely matches the DFT-PAW8 reference, whereas MatterSim exhibits a pronounced tail of attractive radial forces that drive O atoms toward each other. Thus, the unphysical aggregation in MatterSim arises from its overestimation of the attractive tendency between O atoms.

In addition, the previous assessment of the 28 partial PCFs of a chemically complex seven-element iron melt, $Fe_{150}Ni_{15}Si_{25}S_{15}O_{80}C_{15}H_{30}$ , showed deviations in the MatterSim results for many atomic pairs. We find that ECAM achieves nearly perfect agreement with the AIMD results for all partial PCFs (Supplementary Fig. S3).

## 2. Density and seismic velocity

The density and compressional-wave velocity of Fe have been key quantities for characterizing the core. We first assess ECAM's performance in describing the density of pure liquid Fe at 6000 K by comparing its results with previous data in Fig. 3(a). The ECAM-PAW16 results agree well with the AIMD results of Li et al.[47] and Ichikawa et al.[48], as well as the previous deep potential (DP) simulations[38]. The data from Umemoto et al.[20] show systematically higher densities but a similar pressure dependence. The ECAM density is ~2% higher than the value reported by Badro et al. [12] at 6300 K, which lies below all other datasets. The computed densities also agree with the experimental measurements of Anderson et al.[49] and Gan et al.[15]. Although they are systematically higher than those reported by Kuwayama et al.[50], the differences remain within the reported experimental uncertainty.

The compressional-wave velocity ($V_P$) of liquid Fe is computed with ECAM-PAW16 using both the thermal EOS and stress-fluctuation methods. The two methods give consistent results in Fig. 3(b). The previous *ab initio* results compared here were obtained using the thermal EOS method because of the limited cell sizes accessible to AIMD. Our results agree well with those of Umemoto et al.[20] and are slightly lower than those of Ichikawa et al. [48] and Badro et al. [12]. The ECAM velocities also agree well with the velocity estimate based on recent shock measurements at core pressures [15]. Although this estimate corresponds to 5400 K, the weak temperature dependence of $V_P$ in liquid Fe at core pressures supports the comparison [48]. The ECAM velocities are 1–2% higher than those derived from the experimentally constrained EOS of Anderson et al. [49] but agree within the reported uncertainties. They are systematically higher than those of Kuwayama et al. [50]. This difference may partly reflect uncertainties in thermal extrapolation[15]. These comparisons suggest that ECAM accurately reproduces the density and compressional-wave velocity of pure Fe under core conditions.

We next compute the density and compressional-wave velocity ($V_P$) of binary Fe–X alloys (X = Ni, S, O, Si, C, or H) using ECAM-PAW16 at 330 GPa and 6000 K. The effects of Ni and light elements on these key properties have been used to constrain the composition of Earth's outer core through comparison with seismic observations[9,12,20]. To account for differences in pure-Fe reference among studies and isolate the contribution of each alloying element, we compare the relative density and velocity changes as functions of solute concentration. The ECAM calculations in Fig. 4 show that H, C, O, Si, and S reduce the density and increase $V_P$, whereas Ni slightly increases the density and reduces $V_P$. The density varies linearly with concentration, whereas $V_P$ shows a more pronounced nonlinear dependence in Fe–O and Fe–S liquids. The thermal EOS and stress-fluctuation methods yield consistent concentration-dependent $V_P$ values across all six binary systems.

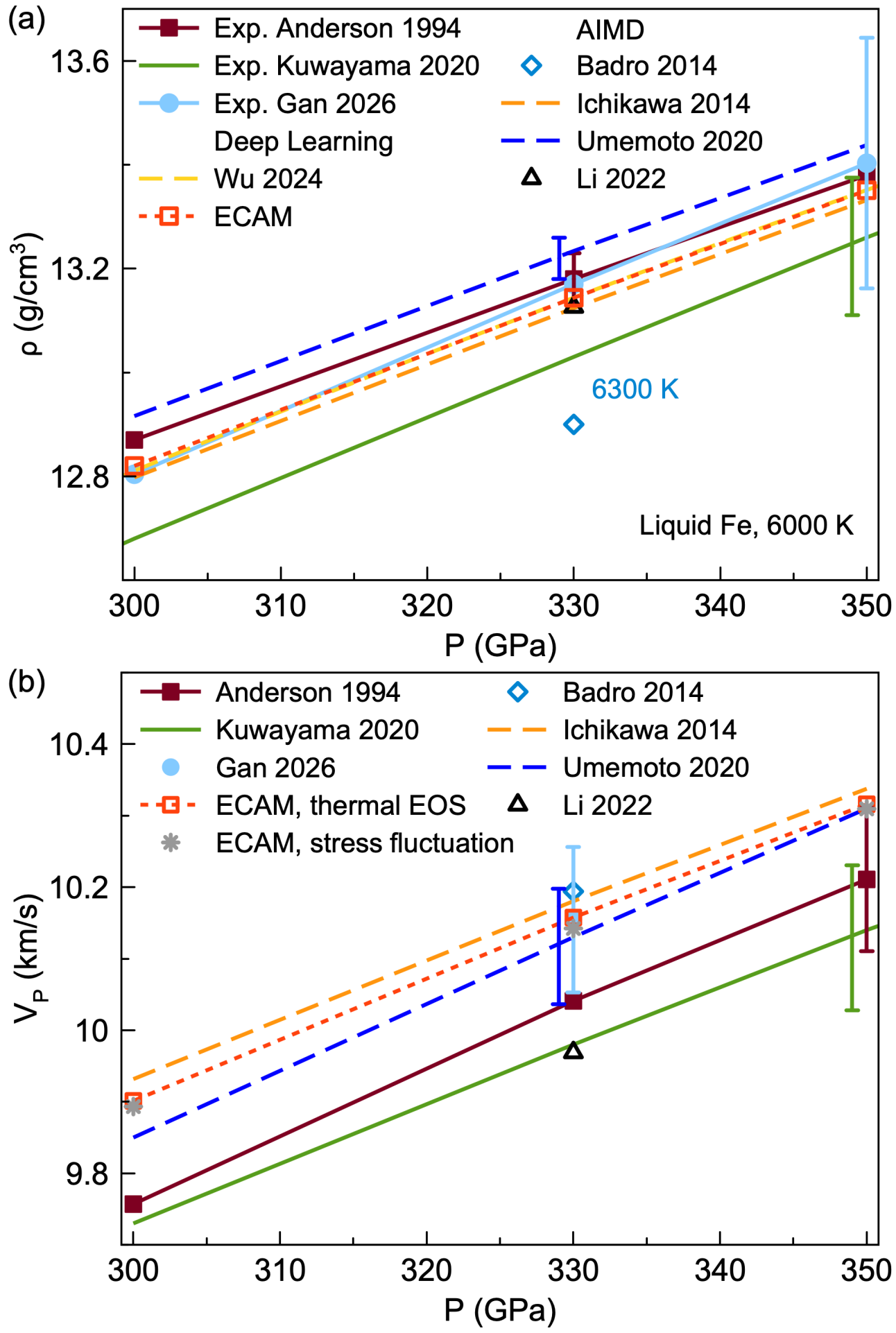

**FIG. 3 Density and compressional-wave velocity of Fe at core pressures.** (a) Density and (b) compressional-wave velocity of liquid Fe at 6000 K. Velocities calculated with ECAM-PAW16 were obtained using both the thermal EOS and stress-fluctuation methods. The data are compared with experimental measurements[15,49,50] and previous calculations [12,20,38,47,48]. The results of Badro et al.[12] correspond to 6300 K.

The ECAM results are compared with previous *ab initio* calculations by Badro et al.[12], Umemoto et al.[20], and Ichikawa et al.[51]. For the latter two studies, we interpolate between the pure-Fe results and those at the lowest reported solute concentrations (ranging between ~20% to 44 at.%) for comparison with our dilute compositions. For Fe–Ni, Fe–S, and Fe–C, both the ECAM densities and $V_P$ values agree well with previous *ab initio* results[12,20]. For Fe–Si, the ECAM $V_P$ values are approximately 0.5% lower than the reference values but remain within the uncertainty range reported by Umemoto et al.[20]. For Fe–O, the ECAM $V_P$ values agree better with those of Badro et al. [12] than with the interpolated results of Ichikawa et al.[51] and Umemoto et al.[20] at low O concentrations. Agreement with the latter datasets improves near 16 at.% O.

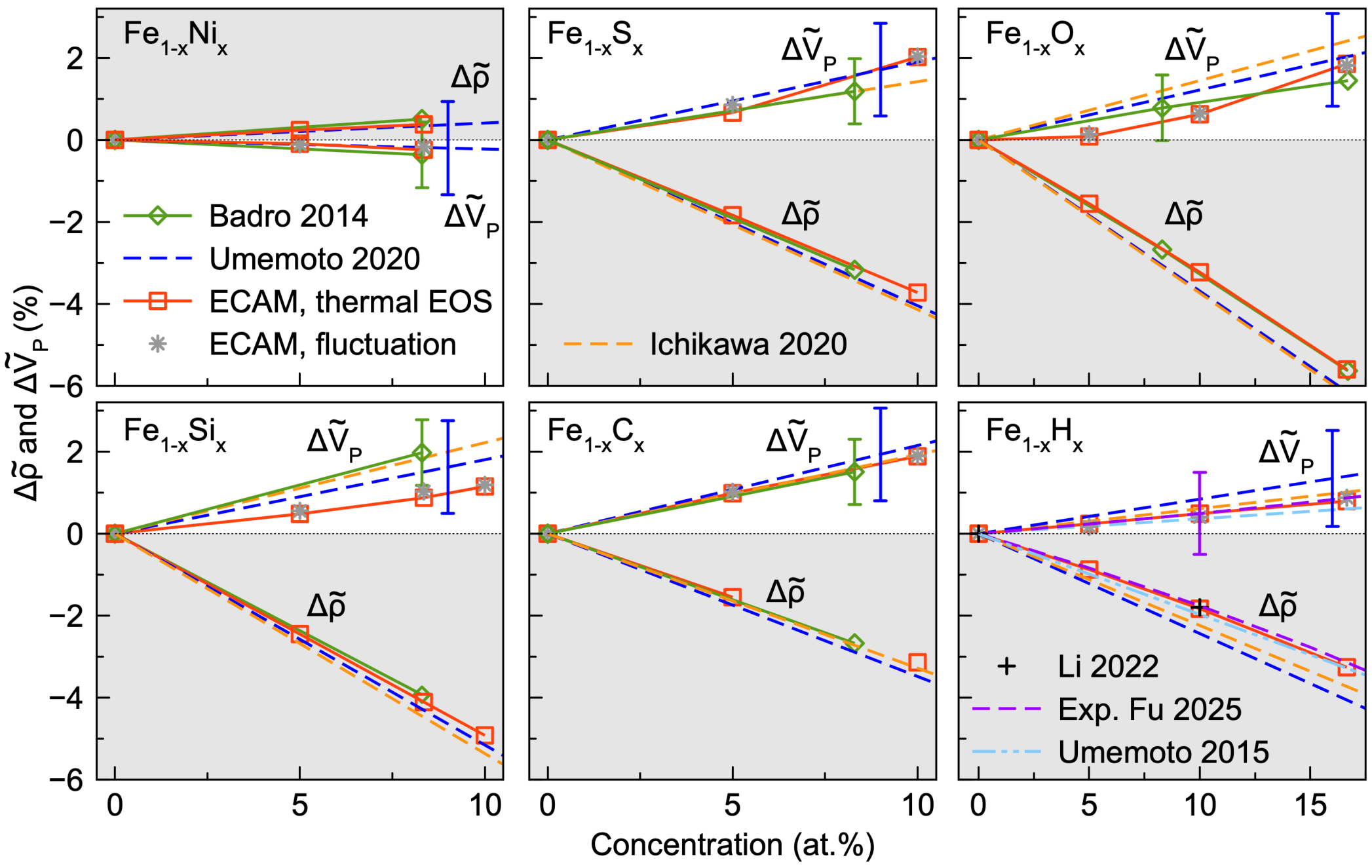

**FIG. 4 Compositional effects on the density and compressional-wave velocity of Fe alloys.** Relative changes in density and compressional-wave velocity of Fe–Ni, Fe–S, Fe–O, Fe–Si, Fe–C, and Fe–H alloys as functions of solute concentration at 330 GPa and 6000 K. Previous computational[12,20,51,53–55] and experimental[52] results are included for comparison, with error bars indicating the reported uncertainties. The results of Badro et al.[12] correspond to 330 GPa and 6300 K. The relative changes are defined as $\Delta\tilde{\rho} = (\rho - \rho^{\mathrm{Fe}})/\rho^{\mathrm{Fe}} \times 100\%$ and $\Delta\tilde{V}_{\mathrm{P}} = \left(V_{\mathrm{P}} - V_{\mathrm{P}}^{\mathrm{Fe}}\right)/V_{\mathrm{P}}^{\mathrm{Fe}} \times 100\%$, where $\rho^{\mathrm{Fe}}$ and $V_{\mathrm{P}}^{\mathrm{Fe}}$ are the corresponding pure-Fe reference values in each study. Red open squares denote ECAM velocity changes obtained using the thermal EOS method, while gray asterisks denote those obtained using the stress-fluctuation method. Gray shading marks the region showing density changes in each panel.

For Fe–H, the ECAM densities and $V_P$ values agree well with recent experimental data from Fu et al.[52]. However, the ECAM densities differ from the interpolated results of Ichikawa et al.[51] and Umemoto et al.[20], whose minimum H concentrations were 26 at.% and 44 at.%, respectively. We therefore also interpolate the Fe–H data of Umemoto et al.[53], which extend down to 22 at.% H. The resulting densities agree better with ECAM. The ECAM $V_P$ values generally agree with previous calculations within the reported uncertainties.

### 3. Gibbs free energy and phase diagram

We further assess ECAM using the Gibbs free energies and phase diagrams of Fe and Fe–H under core conditions and compare the results with recent *ab initio* calculations. Figure 5(a) compares the ECAM free energies of liquid, bcc, and hcp Fe at 323 GPa with previous DFT-PAW16 results[54]. The bcc and hcp free energies from ECAM are slightly lower than the DFT values, whereas the liquid free-energy deviation changes from positive to negative with increasing temperature. The previous DFT calculations have an uncertainty of ~3 meV/atom[54]. Although the ECAM results are well converged with respect to simulation-cell size and trajectory length, the model's training error also contributes to uncertainty in the calculated free energies. Thus, we use the energy RMSE of 4.9 meV/atom as the uncertainty for the ECAM free energies in this comparison. On this basis, the ECAM and DFT results agree within the combined uncertainty ranges. From the intersections of the liquid and solid free energies, we obtain hcp and bcc melting temperatures of 6386(24) K and 6160(50) K at 323 GPa, respectively, consistent with the DFT values of 6357(45) K and 6168(80) K[54]. ECAM thus yields melting temperatures closer to DFT than the MatterSim values of 7045 K for hcp and 6100 K for bcc at the same pressure[31].

Figure 5(b) compares the ECAM-PAW16 melting curve of hcp Fe with previous calculations. The ECAM results closely follow the PAW16 DFT results within the reported uncertainties[54]. Its pressure dependence is consistent with that computed using the DP models of Li et al.[35] and Wu et al.[38], although ECAM computes slightly higher melting temperatures. The two DP melting curves[35,38] agree well with each other despite being based on different DFT codes and pseudopotentials. These models are all based on consistent DFT settings with the same number of effective valence electrons, simulation cells containing more than 250 atoms, and 2×2×2 *k*-point meshes. These curves lie above the estimates of Alfè et al.[55] and Sun et al.[56] and below the estimate of González-Cataldo et al.[24] at 330 GPa, with differences of ~100 K, which are close to the typical error caused by the limited simulation sizes and trajectory lengths in DFT calculations[56].

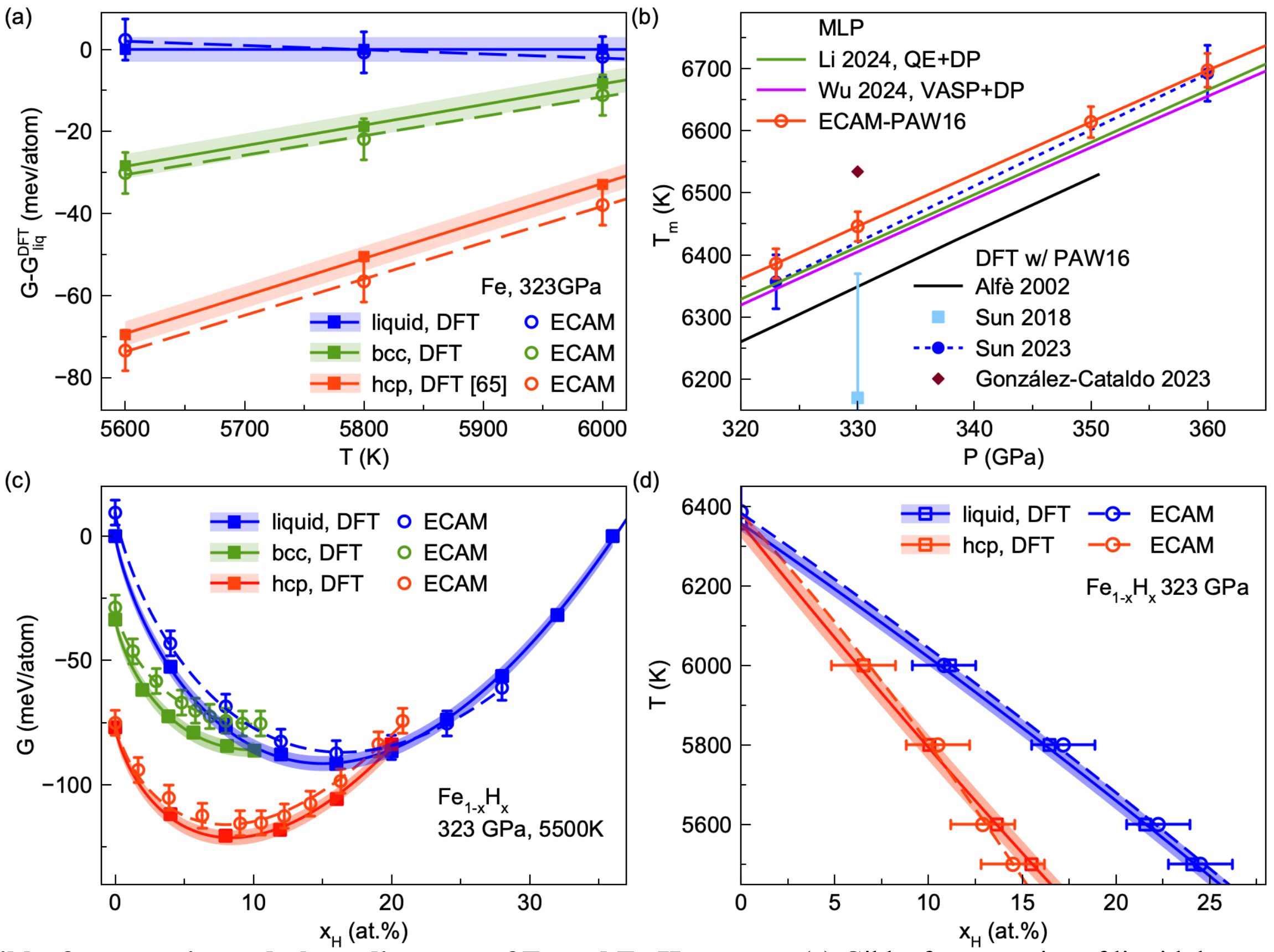

**FIG. 5 Gibbs free energies and phase diagrams of Fe and Fe-H systems.** (a) Gibbs free energies of liquid, bcc, and hcp Fe at 323 GPa, compared with PAW16 DFT results from Ref.[54]. All free energies are referenced to the DFT liquid at the same temperature. Shaded bands indicate the DFT free-energy uncertainties[54], while error bars on the ECAM results are based on the model's energy training error of 4.9 meV/atom. (b) Melting curve of Fe calculated using ECAM-PAW16, compared with previous DFT calculations[24,54–56] and DP models[35,38]. (c) Gibbs free energies of mixing for liquid and superionic bcc and hcp Fe–H at 323 GPa and 5500 K, compared with DFT results from Ref.[27]. All values are referenced to the DFT liquid free energies at 0 and 36 at.% H. (d) Fe–H phase diagram at 323 GPa, compared with Ref.[27].

We further examine ECAM's description of superionic and liquid Fe–H systems. Figure 5(c) compares the ECAM and DFT results[27] at 323 GPa and 5500 K. All free energies are referenced to the linear baseline connecting the DFT liquid free energies at 0 and 36 at.% H. ECAM reproduces the composition dependence and relative free energies of the liquid, bcc, and hcp phases. The small deviations from DFT at intermediate H concentrations are comparable to the error bars. Using the calculated free energies, we construct the Fe–H phase diagram at 323 GPa. Superionic bcc remains metastable over the conditions investigated, so the phase diagram contains only hcp and liquid phases. The ECAM solidus and liquidus agree well with the DFT phase boundaries in Fig. 5(d).

## III. Discussion

By developing ECAM, we illustrate a route to building a domain-specific model for Earth's core materials under extreme conditions using a foundational model with known deficiencies. Although MatterSim provides broad configurational sampling, errors in its local force predictions produce artificial segregation in O-poor liquids. Targeted *ab initio* simulations efficiently supply configurations that MatterSim describes poorly, while active learning with iteratively trained ECAM models extends sampling to environments that MatterSim cannot reliably explore. These additional configurations enable ECAM to eliminate the artifacts present in the parent model. Tests of liquid density and compressional-wave velocity provide complementary validation of ECAM's equation of state and its derivatives, while the phase diagram tests its description of phase competition involving small free-energy differences. Agreement with previous *ab initio* results supports future ECAM studies of complex multicomponent alloys under core conditions.

The final dataset is condensed from nearly half a million MD configurations spanning different compositions in solid, superionic, and liquid phases over a broad range of pressure–temperature conditions relevant to Earth's core. The compact dataset makes it practical to train a model with a lightweight, computationally efficient neural network. As shown in Supplementary Fig. S4, ECAM is almost two orders of magnitude faster than MatterSim, making simulations of kinetic processes such as solid–liquid interface migration during inner-core growth more accessible. Further model compression[57] may bring its computational efficiency closer to that of EAM potentials, potentially supporting simulations of inner-core nucleation[58–60]. Compared with MatterSim, ECAM also explicitly accounts for electronic entropy contributions, which are important for describing core materials at high temperatures[61]. We note that none of the binary systems used in the density, seismic-velocity, and Fe–H phase-diagram validation were explicitly included in the training dataset (Supplementary Table S1). This suggests that the multicomponent training configurations contain sufficient information to describe these simpler binary systems. ECAM should therefore also describe ternary, quaternary, and other combinations of the seven elements under core conditions.

## IV. Conclusion

In summary, by combining foundational-model sampling with targeted *ab initio* calculations, we have developed ECAM, an electronic-temperature-dependent machine-learning model for Fe–Ni–S–Si–O–C–H alloys at 300–400 GPa and 4000–7000 K, covering conditions near the base of Earth's outer core and throughout the inner core. ECAM corrects the artificial O aggregation produced by the foundational MatterSim model and reproduces the density and seismic velocity of liquid alloys, as well as solid–liquid phase equilibria, in general agreement with previous *ab initio* data. Its lightweight neural network enables efficient simulations of solid, superionic, and liquid phases, with a speedup of almost two orders of magnitude over MatterSim. We expect ECAM to open the door to more realistic simulations of Earth's core materials, connecting complex alloy chemistry to seismic observables and core dynamics.

## V. Methods

### 1. First-principles calculations

Density functional theory (DFT) calculations were performed using the Vienna *Ab initio* Simulation Package (VASP)[62] with the projector-augmented wave (PAW) method[63] and the PBE exchange–correlation functional[64]. Two different settings were used in this work. In the PAW8 setting, Fe and Ni were represented with $3d^7 4s^1$ and $3d^9 4s^1$ valence electrons, respectively. The plane-wave cutoff was 400 eV and only Γ point was sampled. In the PAW16 setting, Fe included 16 valence electrons ($3s^2 3p^6 3d^7 4s^1$) and Ni included 18 valence electrons ($3s^2 3p^6 3d^{10}$). The plane-wave cutoff was 750 eV and a $2 \times 2 \times 2$ Monkhorst–Pack mesh was used. In both settings, the valence configurations were $2s^2 2p^4$ for O, $1s^1$ for H, $3s^2 3p^4$ for S, $3s^2 3p^2$ for Si, and $2s^2 2p^2$ for C. Electronic entropy was included through the Mermin functional[65,66]. With the same CPU resources, calculations using the PAW16 settings typically require more than ten times the computational time of those using the PAW8 settings.

### 2. ECAM training

The ECAMs were trained using DeepMD-kit with the smooth-edition se_e2_a descriptor, which encodes both radial and angular information within the local neighborhood of each atom[67,68]. This descriptor was tested for ECAM to provide a sufficiently accurate description of the seven elements. Both the descriptor and electronic temperature were supplied to the fitting network to describe the free-energy contribution [38,69]. Including electronic temperature as an extra model input parameter enables the model to describe the temperature-dependent free-energy surface rather than a fixed ground-state potential-energy surface. This treatment is important for iron alloys under Earth's core conditions, where electronic thermal excitations make a non-negligible contribution to the free energy[38,61].

The embedding network was (25, 50, 100) and the fitting network was (240, 240, 240). The neighbor atom cutoff radius was set to 6.0 Å. The network parameters were randomly initialized and optimized for $2 \times 10^6$ steps using the Adam algorithm[70], with the learning rate exponentially decaying from $1 \times 10^{-3}$ to $3.51 \times 10^{-8}$ during the training process. The loss function was defined as a combination of energy, force, and virial residuals[67]. During training, the energy prefactor was increased from 0.02 to 1, the virial prefactor was increased from 0.02 to 1, and the force prefactor was decreased from 1000 to 1. This schedule emphasizes force accuracy during the early stages of optimization while progressively balancing the energy and virial contributions.

To efficiently capture the structural diversity of a large number of configurations and reduce redundancy, we combined smooth overlap of atomic positions (SOAP) descriptors[71] with principal component analysis (PCA)[72]. For each configuration, local SOAP fingerprints were averaged over atoms to obtain a global structural descriptor. We retained the PCA transformation obtained from the initial dataset throughout all subsequent sampling and data analysis. The PCA map constructed by these reference axes was used to reduce the dimensionality of descriptors, providing a compact representation for the dataset. The PCA map was divided into a uniform grid to select a representative configuration from each cell. In this way, the number of configurations in the final training dataset was reduced by two orders of magnitude.

### 3. Molecular dynamics

Molecular dynamics (MD) simulations using the MatterSim-v1.0.0-5M model [43] were performed with the Atomic Simulation Environment (ASE) [73]. *Ab initio* MD (AIMD) simulations were performed using VASP [74]. MD simulations with the ECAM models were performed using DeepMD-kit and LAMMPS [68,75]. All MD simulations used a time step of 0.5 fs and the Nosé–Hoover thermostat.

**4. Seismic velocity calculations**

The compressional-wave velocity was calculated as $V_P = \sqrt{K_S/\rho}$, where $\rho$ is the density and $K_S$ is the adiabatic bulk modulus. We computed the adiabatic bulk modulus using both the conventional thermal equation-of-state (EOS) and stress-fluctuation[76,77] methods.

In the thermal EOS method, to obtain $K_S$ at a specific volume $V_0$, we first performed a series of NVT simulations at a fixed temperature and several volumes around $V_0$. The pressure–volume data were fitted to the third-order Birch–Murnaghan equation of state to compute the isothermal bulk modulus $K_T = -V(\partial P/\partial V)_T$ at $V_0$. The adiabatic bulk modulus was computed using

$$K_S = K_T(1 + \alpha\gamma T). \tag{1}$$

where $\alpha$ and $\gamma$ are the thermal expansion coefficient and Grüneisen parameter, respectively. The thermal expansion coefficient $\alpha = V^{-1}(\partial V/\partial T)_P$ was computed from a series of NPT simulations at several temperatures at the pressure $P$ corresponding to $V_0$. Determining the Grüneisen parameter requires the relationship between pressure and internal energy at fixed volume. We therefore performed additional NVT simulations at different temperatures at $V_0$, setting the electronic temperatures $T_{el}$ equal to ionic temperatures $T_{ion}$ at each state point. ECAM directly outputs $F_{ECAM} = E_p - T_{el}S_{el}$ without explicitly providing the electronic entropy $S_{el}$. The potential energy $E_p$ therefore needs to be computed separately to obtain the internal energy $U$. To this end, we held the configurations from an NVT trajectory fixed and recalculated $F_{ECAM}(T_{el})$ at different electronic temperatures $T_{el}$ using the RECAL protocol [78]. The electronic entropy $S_{el}$ was then computed by differentiating $F_{ECAM}(T_{el})$ as $S_{\mathrm{el}} = -\frac{\partial F_{ECAM}(T_{el})}{\partial T_{el}}$. The internal energy $U$ was obtained as

$$U = K_{\mathrm{ion}} + F_{ECAM} + T_{el}S_{el}, \tag{2}$$

where $K_{\mathrm{ion}}$ is the ionic kinetic energy. We then obtained $\gamma = V(\partial P/\partial U)_V$ by fitting the mean pressure against the internal energy $U$ along the fixed-volume temperature series.

For the stress-fluctuation method, we first performed NPT simulations to determine the equilibrium volume and density at each pressure–temperature condition. The resulting cell was equilibrated in the NVT ensemble for 100 ps, followed by a 200 ps adiabatic sampling run in the NVE ensemble at fixed electronic temperature. The adiabatic thermodynamic elastic tensor $A_{ijkl}$ at $T$ was calculated as[77]

$$A_{ijkl} = \langle A^B_{ijkl}\rangle - \frac{V}{\mathrm{k_B T}}\left[\langle\sigma_{ij}\sigma_{kl}\rangle - \langle\sigma_{ij}\rangle\langle\sigma_{kl}\rangle\right] + \frac{2Nk_BT}{V}\left(\delta_{il}\delta_{jk} + \delta_{ik}\delta_{jl}\right), \tag{3}$$

where $\langle\cdot\rangle$ denotes the ensemble average over the simulation time. $V$ is the simulation-cell volume, $N$ is the total number of atoms, and $k_B$ is the Boltzmann constant. The stress $\sigma_{ij}$ and Born matrix $A^B_{ijkl}$ characterize how the energy changes under an infinitesimal deformation $\epsilon_{ij}$. The former is the total instantaneous stress, including both virial and kinetic contributions, and equals the negative of the pressure tensor $P_{ij}$ from MD. The latter is given by the second-order derivative $A^B_{ijkl} = \frac{1}{V}\frac{\partial^2 F_{ECAM}}{\partial\epsilon_{ij}\partial\epsilon_{kl}}$. Here, $\sigma_{ij}$ and $A^B_{ijkl}$ were sampled every 20 and 200 MD steps, respectively. At finite pressure, the thermodynamic elastic tensor $A_{ijkl}$ was converted to the effective stiffness tensor $C_{ijkl}$ using the Barron-Klein correction,

$$C_{ijkl} = A_{ijkl} + P\left(\delta_{ij}\delta_{kl} - \delta_{il}\delta_{kj} - \delta_{ik}\delta_{jl}\right), \tag{4}$$

where $P$ denotes the pressure $P = -\frac{1}{3}\mathrm{Tr}\langle\sigma\rangle$ when the stress is hydrostatic. After an electronic entropy correction was applied to $C_{ijkl}$, the adiabatic bulk modulus $K_S$ was obtained using

$$K_S = \frac{[(c_{11} + c_{22} + c_{33}) + 2(c_{12} + c_{23} + c_{31})]}{9}. \tag{5}$$

**5. Free energy calculations**

We calculated the Gibbs free energies using Hamiltonian thermodynamic integration (TI) from embedded-atom method (EAM) reference systems to ECAM [54]. The EAM potentials and corresponding reference free energies $G_{\mathrm{EAM}}$ for the pure Fe and Fe–H systems were taken from [54] and [27], respectively. For phase $\alpha$, the ECAM Gibbs free energy was computed as

$$G^\alpha_{\mathrm{ECAM}}(T) = F^\alpha_{PV}(T) + F^\alpha_{TI}(T) + G^\alpha_{\mathrm{EAM}}(T) \tag{6}$$

where $\alpha$ labels the superionic, solid and liquid phases. The volume term is

$$F^\alpha_{PV}(T) = P(V^\alpha_{ECAM} - V^\alpha_{\mathrm{EAM}}) - \left(\int_{V^\alpha_{\mathrm{EAM}}}^{V^\alpha_{ECAM}} P^\alpha_{\mathrm{EAM}}(V)dV\right) \tag{7}$$

Here, $V^\alpha_{ECAM}$ and $V^\alpha_{\mathrm{EAM}}$ are the equilibrium volumes of phase $\alpha$ at given $(P, T)$ for ECAM and EAM, respectively. The ECAM equilibrium volumes were obtained from NPT simulations. $P^\alpha_{\mathrm{EAM}}(V)$ is the EOS of phase $\alpha$ for EAM. The TI term $F^\alpha_{TI}(T)$, which accounts for the Helmholtz free-energy difference between ECAM and EAM, was computed using a nonequilibrium TI scheme implemented in LAMMPS [79]. In the TI simulations, the force acting on each atom was given by $f = (1-\lambda)f_{\mathrm{EAM}} + \lambda f_{\mathrm{ECAM}}$. A Nosé–Hoover thermostat was used to maintain the target temperature, and a time step of 0.5 *fs* was used to integrate Newton's equations of motion in the TI-MD simulations.

**Acknowledgments**

Work at Xiamen University was supported by the National Natural Science Foundation of China (Grant No. 42550120 and T2422016). R.M.W acknowledges support from the Gordon and Betty Moore Foundation (GBMF12801). The Tan Kah Kee Supercomputing Center are acknowledged for their support in GPU computing.